\documentclass[twoside, 10pt]{article}
\usepackage{zif,epsfig}

\begin{document}
\setlength{\textheight}{7.05truein}    

\runninghead{Running Title}
            {Running Author(s)}

\normalsize\textlineskip \thispagestyle{empty}
\setcounter{page}{489}

\copyrightheading{}{}{2005}{489--498}

\vspace*{0.88truein}

\alphfootnote

\fpage{197}

\centerline{\bf
QUANTUM MEASUREMENT ACT} \vspace*{0.035truein}
\centerline{\bf AS A SPEECH-ACT}
\vspace*{0.37truein} \centerline{\normalsize
JEAN SCHNEIDER}
\vspace*{0.015truein}
\centerline{\small\it CNRS - Paris Observatory}
\baselineskip=11pt
\centerline{\small\it 92190 Meudon, France}
\baselineskip=11pt
\centerline{\small\it (jean.schneider@obspm.fr)}

\vspace*{0.225truein}

\vspace*{0.21truein}

\abstracts{{\bf Abstract:}
I show that the quantum measurement problem can be understood if the
measurement is seen as a ``speech act'' in the sense 
of modern language theory. The 
reduction
of the state vector is in this perspective
an intersubjectice -- or better a-subjective --
symbolic process. I then give some perspectives on 
applications to the ``Mind-Body problem''. }{}{}

\vspace*{10pt}


\vspace*{1pt}\textlineskip    

\section{Introduction: ``Realism'' is an Idealism}

Science, and in particular physics, is a perpetual fight against absolute 
basements
and essences:
space, simultaneity, heat as phlogistic, ether, properties of quantum objects.
Relativity theory and quantum physics have shown all the benefits of the 
renouncement of essences such as ether and values of physical quantities.
Such essences do not belong to experience and are only the fruit of imagination
(more exactly they represent an abstraction constructed out of experience 
thanks to {\it a priori} concepts).
In this sense, if by ``realism'' one means the belief that there is an 
essence behind experience, realism is an idealism. An ``objective underlying 
reality'' is only a word (expressing a desire of reality) 
 and there is nothing behind or beyond it. Since the present meeting is also 
devoted to ``the subjective'', it is worthwhile to point out that the same 
phenomenological and constructivist approach holds also for the mental 
world and that for instance
``the'' mind (as an essence) has also to be renounced.

\section{The quantum measurement}
\subsection{Reminder of the problem}
Although the rules of quamtum mechanics are well known, it is better, for 
clarity, to recall them.
They rest on primitive notions such as "system",
 "state of a system", "observable" and can be summarized as follows:
\begin{itemize}
\item
R1   Every system S is described by a Hilbert space $Hilb$.
\item
R2   Any state of the system is described by a $\psi \in Hilb$
\item
R3   In absence of measurement, the  system evolves according to the 
Schr\"odinger equation 
      $i \hbar \partial \psi /\partial t =  H\psi$, where $H$ is the 
hamiltonian of      the system
\item
R4   A physical quantity (observable) is described by an operator
$A$  on $Hilb$
\item
R5   The only possible outcomes of a measurement of the observable represented 
by  $A$ are 
        the proper values $a_i$ of $A$, with the corresponding proper vectors:
        $A\psi _i = a_i\psi _i$.
\item
R6   The  result of the measurement of $A$ on the system
in a state $\psi$ is random with a probability given
by $p_i =
        |<\psi _i|\psi>|^2$
\item
R7   After the measurement the system is in the state $\psi _i$
(``state vector collapse'').
\end{itemize}
There is a kind of duality in these fundamental concepts and rules, since rules 
R1 - R3 deal with  the description of the system, while rules R4 - R6 deal with
observables which appear heterogeneous with respect to the system. In this 
sense, the observables do not belong to the system.

It is natural for a physicist to try to describe
the measurement as an interaction between the system and the apparatus 
and therefore the latter as an other system, i.e. by a state vector 
$\psi _A$ of 
$Hilb$.
But then, when this approach is translated into the quantum formalism, 
     a contradiction appears. Indeed, let  $\psi_{SA}$ be
the vector describing the meta-system "system + apparatus" and $H_{SA}$ the 
interaction operator system-apparatus.
Then:
\begin{itemize}
\item
from (R3), after the measurement, the meta-system is in the 
     (unique and predictable) state $\psi_{SA}(t)=e^{-i/\hbar H_{SA}t}
\psi_{SA}(0)$.
\item
from (R7), after the measurement, the system is, at random, in one of the 
states $\psi _i$. 
\end{itemize}
The two final states are incompatible. That is the problem.

The central question then is: ``Why does the process of observation
(giving rise to the state vector collapse, that is to a sudden transition
between two states of the observed system) escape  the normal evolution of
the pair system$+$observer described by the Schr\"{o}dinger equation?''. 
There is an even more radical question. The knowledge of the 
state $|\psi>$ of the system is necessary to predict the possible outcomes of
the observation. But it is not sufficient since, to describe the set of 
outcomes,
we need to add an heterogeneous element, the operator associated with the
observable which is measured. Why is this second level necessary ? I shall
call it the ``question of the concept of observable''.

\subsection{Why ``decoherence'' does not solve the problem.}
Several solutions have been proposed during the past years. Some of them 
modify in a way or another the foundations of quantum mechanics: hidden  
variables, spontaneous localization, non linear Schr\"{o}dinger equation,
``many worlds'' (in fact many observers) interpretation etc.

A different solution, known as the 
decoherence theory, has been developed
by Zeh, Zurek, Omn\`es and others without any change in the standard 
postulates. It consists in pointing 
out that 
the interaction of the system with the environment diagonalizes very rapidly, 
with a very short characteristic time $\tau$ and in an irreversible manner, the density
matrix of the meta-system formed by the system, the observer and the
environment, thus leading to an apparent quasi-collapse of the
state vector. This explanation has become popular since the occurence of
decoherence has been experimentally demonstrated  (Davidovich, Brune,  Raimond 
et al., 1996).

Unfortunately the explanation based on decoherence is not satisfying for the 
following 
reasons. First, decoherence is a statistical notion based on the statistical
matrix representing statistical ensembles of systems. But in a given 
experiment one does not deal with statistical ensembles but with an individual
system (and an apparatus). In other words, the unicity of the result of a 
given experiment is not expressed by a diagonal matrix. In mathematical terms,
decoherence leads to a diagonalized matrix, while in a single experiment
all the diagonal elements of the matrix are all zero except one. In other words, 
the expression 
\begin{center}
``$\rho =\left(\begin{array}{cc}1/2&0\\0&1/2\end{array}\right)$'' 
\end{center}
is not
the same as 
\begin{center}
``$\rho =\left(\begin{array}{cc}1&0\\0&0\end{array}\right)$ ~
or ~ $\rho =\left(\begin{array}{cc}0&0\\0&1\end{array}\right)$''.
\end{center}
Supporters of decoherence often reply that quantum
physics makes only statistical predictions. That statement is contradicted
by predictions like ``the measurement of any component of
the spin of a photon in a single
experiment will give an integer result''. In addition, the very concept of
``ensemble'' presupposes that there are single individuals of the ensemble.
If quantum theory only deals with statistical predictions, it is an
incomplete theory since individual experiments escape it.
A second problem with decohrence refers to the  question of the concept of 
observable (end of section 2.1). This question is in fact 
addressed to any attempt to describe
the measurement as a system-apparatus interaction. If this interaction was a
good model, it should be able to describe several aspects of a measurement
with, as only primitive concepts, those of state vectors $\psi_S$, $\psi_A$
of the system and the apparatus 
 and the system-apparatus interaction hamiltonian. Namely:
\begin{itemize}
\item What is the meaning of an ``observable''?
\item
 What means ``the value'' of an observable?
\item
 Why is the outcome of an experiment random, while the interaction is 
  deterministic?
\item
 After the interaction why is the system precisely in one of the states
  $|a_i>$?
\item
Why are the only possible outcomes one of the proper values of an operator
  $A$ (that the model should construct)?
\end{itemize}
J.-S. Bell was well aware of all these difficulties when he wrote his paper 
``Against measurement'' (1990) where he proposed to replace observables by 
``beables''.

\subsection{What is really a measurement?}
To be performed, a measurement needs two ingredients:
\begin{itemize}
\item an apparatus, object of perceptions and manipulations
\item
 (pre-existing) mathematical symbols to express the result.
\end{itemize}
There is indeed no measurement without (or before) 
the expression (in mathematical terms, for instance ``$A = a_i$'') of its 
result. This is not a 
philosophical point of view, it is an empirical fact. In this respect, 
$A = a_i$
 \underline{does not}  reflect or translate a reality outside  itself.
    It creates, by its own declaration, this reality. It \underline{is} 
this (symbolic) reality.  As a matter of fact, a symbol is
its own actualization. It means that, as a mathematical symbol, the outcome 
of a measurement is not the (quantum) state 
of the screen of an apparatus and, thus, cannot be described by a state vector.
In pre-quantal terms, a symbol, {\it e.g.}
 {\bf 1}, is differrent from its pixelized 
image and from its 
physical support, since the symbol {\bf 1} is required {\it a priori} 
before any 
pixelisation (Figure 1). In terms of interaction, a measurement is thus not 
a physical interaction (i.e. 
described 
by an Hamiltonian) between two systems (described by state vectors), but 
an ``interaction''
between language (discourse) and a perception.

\begin{figure} [ht]
\centerline{\epsfig{file=pixel.epsi, width=7.50cm}} 
\vspace*{12pt}
\fcaption{\label{motion}The pixelisation of the symbol `` {\bf 1} `` 
is different from the symbol itself.}
\end{figure}

  These remarks lead in a natural way to the solution I 
propose (Schneider 1994):
 the measurement 
act is not a physical transition or phenomenon, but a purely 
\underline{semantic} act, in the same line as the {\it speech acts}
\footnote{For
a general introduction to these notions, see J. Austin 1982} .
well known in language theory. A speech act does not describe a situation 
independent of itself, it creates what at the same time it describes. The
measurement act has more precisely the structure of a {\it declaration}. 
The question whether this process is of psychological nature or takes places
in some mind is not relevant. A semantic process is exterior to any individual,
it is existing only as
shared by the community of locutors and in this sense is objective. It just
takes place in a symbolic universe, the universe of discourse in which all
physicists live. It is  the universe studied by linguistics and
semiotics. It has nothing to do with psychology.
It is not the ``consciousness'' of the observer which operates the
state vector collapse, as was proposed by London and Bauer (1983). 
It is the result of an
impersonal, non psychological but empirically ascertainable, production of a
{\it signifier} which exists only as shared by the community of 
physicists \footnote{According to modern views, consciousness is on the
contrary defined as being the crossrads of different {\it signifier}.} .
In other words it is not a {\it passive} registration, it is an {\it active} 
semantic
process.
The subjectivity of one observer is to be replaced by the intersubjectivity of
the discourse, with no psychological subject, where the 
impersonal semantic collapse of the state vector takes place. 
 To express it in another way, the measurement 
act, as giving an attribute to a system, is an act of attribution, a
declarative act. The judicial domain can help us for an analogy: a judgement
does not register afterward a pre-existing reality, it does create it by its 
verdict. The judgement ``guilty'' creates, in the judicial universe, 
guilt. 
 The result of that act is of course random and has
a probability of occurence $|<a_i|\psi >|^2$. This conception sheds a new light
on causality in the quantum measurement: the result of a measurement act has 
no other cause than itself, it is its own cause. It is in this respect that 
there is no quantum causality.\\

The ``classical'' character of the measurement apparatus lies in the semantic 
nature of its description, not in its complex atomic structure (as could
naturally but erroneously be infered from the Ehrenfest theorem). A system is a
measurement apparatus only insofar as it is described by a set of
{\it signifiers}; otherwise
it is nothing but a quantum system. As for the observer, it is most
certainly decomposable in
 atoms, but it is an observer only as a support of semantems.
In a measurement, the so-called interaction with the measuring apparatus
(which would be described by an Hamiltonian) is an encounter, an interaction 
if one may say so, between the observed system and the universe of discourse.
Because this encounter is not descriptible by an Hamiltonian  the 
measurement process escapes the Schr\"{o}dinger equation. It was N. Bohr 
(1983) who was
among the first authors pointing out the role of language in the 
measurement. But for him language was just a collection
of words, the vocabulary of classical physics. Here the point of
view is different:
what is important is not so much the {\it content}, but the 
{\it auto-productive} nature of a {\it signifier}
and it is this auto-production which gives rise to the state vector collapse. 
\\

The idea that a measurement does not result from an interaction between a
system and an apparatus has been also recently been expressed by
O. Ulfbeck and A. Bohr (2001). For these authors, quantum physics does only
deal with clicks of an apparatus. But they do not
address the essential question: "What is a click?". For instance when a 
click is 
recorded in a movie, what is the real click? The click or the movie of the 
click?
The present paper explicitly claims that the objective (intersubjective)
click is the
declaration: "There is/was a click".

We can now apply this constructivist approach to the notion of subject
and to the Mind-Body problem.
\section{Mind-Body}
\subsection{General principles}
We have seen that a measurement  is the (random) emergence of
a symbol detached from (the appearance of) an apparatus. If the symbol is 
mathematical, it is a scientific (physical) measurement. 

But there can be ``pre-scientific'' measurements when the symbol is vague or 
fuzzy such as a colour, a sound, a smale etc. There is then (at least up to
now) no mathematical representation of these vague symbols by hermitian
operators. But they do nevertheless exist (i.e. are experienced) as symbols.
All these vague symbols are not systems and have no state vector.

A first application of  vague symbols in the context of quantum physics
is the answer they provide   to an argument often opposed (in particular
by J.-S. Bell) to the point of view defended here that the state vector 
collapse is operated by an observer. The argument is formulated in ironic
terms: ``Only physicists having their PhD can operate a state vector 
collapse''. In other words, observations or experiments made without the 
support of elaborated mathematics would not exist. The notion of vague
symbol provides an answer: every experience, whatever its vagueness, is 
legitimate at its own level. It is a scientific measurement when it is
expressed in scientific symbols.

Vague symbols lead to a more general notion of symbol, introduced 
progressively along all 
the XX$^{th}$ century by semiotics (the science of symbols). Indeed, words 
and mathematical symbols are special types of symbols. Symbols are what 
Cassirer calls symbolic forms. They are, like Kantian concepts, {\it a priori}
symbols. They belong to an unlimited variety of registers: acoustic, graphical,
gestual, conceptual, judicial, institutional, esthetical, emotional,
affective, ethical etc.
They are all structured as declarations designating what they construct.
To be less elliptic, this structure means that in a first step, as a declarative
gesture, they produce themselves and in a second (timeless) step they
present themselves as designating
 from the outside, as an objective reality, what they 
have just created \footnote{This process leads to the notion of 
``afterwardness'', a non linear notion of time, described by J. Lacan.in his
work ({\it passim})} . To 
illustrate this approach by a concrete example, the symbol `` a `` 
is, in a first time, just a given letter which, in second time, designates 
the notion
of ``symbol a''. It represents a kind of self-distanciation of symbols.

As mentioned in the introduction, no genuine ``consciuosness''
nor ``subject'' is needed . They are not genuine instances, they are 
constructed
objects out of two primitive instances: subject-less
sensations and (declarative) symbols.
This construction, also called symbolization, detaches a symbolic object
from the sensation. To be more precise, there is a primitive instance, the
so called ``object-relation'' (equivalent to a sensation)
 which is a complex made of a relation
and its ``to be'' object, entangled together. 
At this point, it is not relevant to ask if 
the object-relation is one or two instances, since the concept of number does
not apply: we are in the realm of a ``proto-arithematic'' (Schneider 1994). 
More exactly,  symbolization creates attributes  and
an object is in a second step the synthesis of different attributes.

\subsection{Tentative quantum modelization of the Mind-Body relation}
To address this question,  Mind and  Body have first to be defined 
and characterized in the framework of the concepts presented here
(Schneider 1997).

``The'' mind, or the subject, as things are bad primitive concepts. They 
have to be replaced by a-subjective symbols, {\it i.e.} symbols by their own,
source-less. In the present view, the ``subjective'' 
is then a particular object: an object constructed out of ethical symbolic
forms \footnote{The processes by which the subjective is constructed
are very complex, they involve parental and
social discourses, words like ``I'' which precede the subject, identification
etc; rigorously speaking, the sentance ``I speak'' means something like
``The word ``I'' speaks''. That is why the traditional
subjective is in reality a-subjective.} .

The physical body is \underline{not} the source of sensations. As a
physiological object, it is an abstraction constructed by a bio-physical
theoretization out of primitive and source-less sensations.

In other words, the primitive concepts are no more  Mind and  Body,
but sensations and symbols out of which  Mind and  Body are 
constructed abstract objects. In particular, the body is an abstract
synthesis of physiological attributes resulting from symbolization.

In quantum theory, symbolic attributes ({\it i.e.} values of observables)
emerge randomly and are cause-less. By extending the notion of symbol
as in section 3.1, there are two types of bodies created by symbolization
out of sensations:
\begin{itemize}
\item the physical, or physiological body, {\it i.e.} the bio-physical  description
of the body created by the conceptualization of physics
\item
 the emotional body created by emotional symbols (words of pain, joy, 
anxiety etc).
\end{itemize}
Emotional symbols are genuine, not constructible from physiological instances.
This conception is generalizable to non verbal symptoms (I refer here to
the psycho-analytical conception of symptoms as symbols).

Take for instance as physiological observables  skin colour, cardiac 
rhythm, blood pressure.
The emotional observables are for instance an exchange of words (with or 
without an emotional content with an
interlocutor). A complete discussion should include
unconscious aspects, always emotional, of symbols. 
 The two types of observables do not ``commute'', they are 
complementary in the quantum mechanical sense: it means that
an individual cannot at the same time be subject to a physiological
observation and have emotional relationships. It is interesting to note that
C. Bohr (father of N. Bohr and biologist) wrote: 
\begin{center}
\begin{tabular}{l}
``An organism cannot at the 
same time be subject to a chemical \\
analysis and be declared as living''. 
\end{tabular}
\end{center}

We then can have  a succesion of non commutative events to describe
how an emotion can make a face blushing: white skin $\longrightarrow$ 
expression of emotion $\longrightarrow$ pink skin. It is similar to the
 quantum
measuremets of non commutative components of ths spin: $S_X = +1/2$ 
 $\longrightarrow$ $S_Z = +1/2$  $\longrightarrow$ $ S_X = -1/2$.
We so have  a simplified scheme for 
quantum modelization of the undeterministic evolution
of the body.

\section{Perspectives}
The main stream in current cognitive sciences is to seek
a ``naturalization'' of consciousness. It is an attempt to treat
Mind and consciousness as objects (however immaterial they are).
An essential prediction  of the present approach is that
 these attempts of  
 naturalization  will certainly improve our knowledge of the physical brain
but not of the mind.

Secondly, many authors attempt to reconstruct, essentially thanks to
decoherence, the classical world out of the quantum level. In the present 
approach, it is the classical world which precedes the quantum level: the 
latter is constructed from the behavior of macroscopic apparatuses.

With the concept of afterwardness briefly discussed
in section 3.1 (and formalized in Schneider (1994)) it 
becomes possible to
reformulate the notion of consistent history ({\it e.g.} Omn\`es 1994) and
the transform it into a notion of ``afterward history'' (Schneider 2000). \\

\end{document}